\documentclass[12pt,onecolumn,journal]{IEEEtran}
\usepackage[margin=2.0cm, lmargin=2.0cm]{geometry}
\usepackage{latexsym}
\usepackage{amsfonts}
\usepackage{amsbsy}
\usepackage{cite,amsmath,amssymb}
\usepackage{times}
\usepackage{graphicx}
\usepackage{enumerate}
\usepackage[usenames]{color}
\usepackage[dvips]{pstcol}
\usepackage{epstopdf}
\usepackage{caption}
\usepackage[caption=false,font=footnotesize]{subfig}
\input epsf

\title{Modulation and Multiple Access for 5G Networks}

 \author{Yunlong Cai,  Zhijin Qin, Fangyu Cui, Geoffrey Ye Li, and Julie A. McCann
\thanks{

Y. Cai and F. Cui are with the College of Information Science and Electronic Engineering, Zhejiang University, Hangzhou 310027, China (e-mail: ylcai@zju.edu.cn; cfy531@zju.edu.cn).

Z. Qin and J. A. McCann are with the Department of Computing, Imperial College London, London, UK, SW7 2AZ (e-mail: z.qin@imperial.ac.uk; jamm@imperial.ac.uk).

G. Y. Li is with the School of Electrical and Computer Engineering, Georgia Institute of Technology, Atlanta, USA (e-mail: liye@ece.gatech.edu).

 } }

\begin{document}

\maketitle \thispagestyle{empty}

\vspace{-1cm}
\begin{abstract}
Fifth generation (5G) wireless networks face various challenges in order to support large-scale heterogeneous  traffic and users, therefore new  modulation and multiple access (MA) schemes are being developed to meet the changing demands. As this research space is ever increasing, it becomes more important to analyze the various approaches, therefore in this article we present a comprehensive overview of the most promising modulation and MA schemes for 5G networks. We first introduce the different types of modulation that indicate their potential for orthogonal multiple access (OMA) schemes and compare their performance in terms of spectral efficiency, out-of-band leakage, and bit-error rate. We then pay close attention to various types of non-orthogonal multiple access (NOMA) candidates, including power-domain NOMA, code-domain NOMA, and NOMA multiplexing in multiple domains. From this exploration we can identify the opportunities and challenges that will have significant impact on the design of modulation and MA for 5G networks.
\end{abstract}


%
\IEEEpeerreviewmaketitle

\section{Introduction}
In recent years, fifth generation (5G) wireless networks have attracted extensive research interest. According to the 3rd generation partnership project (3GPP)~\cite{3GPP,ITU}, 5G networks should support three major families of applications, including enhanced mobile broadband (eMBB) \cite{3GPP,ITU}; massive machine type communications (mMTC)~\cite{3GPP,ITU}; and ultra-reliable and low-latency communications (URLLC)~\cite{3GPP,ITU}.  On top of this, enhanced vehicle-to-everything (eV2X) communications are also considered as an important service that should be supported by 5G networks~\cite{3GPP}. These  scenarios require massive connectivity with high system throughput and improved spectral efficiency (SE) and impose significant challenges to the design of general 5G networks. In order to meet these  new requirements, new  modulation and multiple access (MA) schemes are being explored.

Orthogonal frequency division multiplexing (OFDM) \cite{OFDM, book, OFDM2} has been adopted  in fourth generation (4G) networks. With an appropriate cyclic prefix (CP), OFDM is able to combat the delay spread of wireless channels with simple detection methods, which makes it a popular solution for current broadband transmission. However, traditional OFDM is unable to meet many new demands required for 5G networks. For example, in the mMTC scenario \cite{3GPP,ITU}, sensor nodes usually transmit different types of data asynchronously in narrow bands while OFDM requires different users to be highly synchronized, otherwise there will be large interference among adjacent subbands.

To address the new challenges that 5G networks are expected to solve,
various types of modulation  have been proposed, such as filtering, pulse shaping, and precoding to reduce the out-of-band (OOB) leakage of OFDM signals. Filtering \cite{UFMC,UFMCvsFBMC2,fofdm,fofdm2} is the most straightforward approach to reduce the OOB leakage and with a properly designed filter, the leakage over the stop-band can be greatly suppressed. Pulse shaping \cite{Farhang2011,PPN,GFDM2,GFDM}  can be regarded as a type of subcarrier-based filtering that reduces overlaps between subcarriers even inside the band of a single user, however, it usually has a long tail in time domain according to the Heisenberg-Gabor uncertainty principle~\cite{uncertainty}.
Introducing precoding \cite{spofdm,ncontiofdm,ncontiofdm2,notchofdm} to transmit data before OFDM modulation is also an effective approach to reduce leakage. In addition to the aforementioned approaches to reduce the leakage of OFDM signals, some new types of modulations have also been proposed specifically for 5G networks.
For example, to deal with high Doppler spread in eV2X scenarios, transmit data can be modulated in the delay-Doppler domain \cite{OTFS}. The above modulations can be used with orthogonal multiple access (OMA) in 5G networks.
%

OMA is core to all previous and current wireless networks; time-division multiple access (TDMA) and frequency-division multiple access (FDMA) are used in  the second generation (2G) systems, code-division multiple access (CDMA) in the third generation (3G) systems, and orthogonal frequency division multiple access (OFDMA) in the 4G systems. For these  systems,  resource blocks are orthogonally divided in time, frequency, or code domains, and therefore there is minimal interference among adjacent blocks and makes signal detection relatively simple. However, OMA can only support limited numbers of users due to limitations in the numbers of orthogonal resources blocks, which limits the SE  and the capacity of current networks.
 To support a massive number of and dramatically different classes of users and applications in 5G networks, various NOMA schemes have been proposed.


As an alternative to OMA, NOMA introduces a new dimension by perform multiplexing within one of the classic time/frequency/code domains. In other words, NOMA can be regarded as an ``add-on'', which has the potential to be harmoniously integrated with existing MA techniques. The core of NOMA is to utilize power and code domains in multiplexing to support more users in the same resource block. There are three major types of NOMA: power-domain NOMA, code-domain NOMA, and NOMA multiplexing in multiple domains.  With NOMA, the limited spectrum resources can be fully utilized to support more users, therefore  the capacity of  5G networks can be improved significantly even though extra interference and additional complexity will be introduced at the receiver.

To address the various challenges of 5G networks, we can either develop novel modulation techniques to reduce multiple user interference for OMA or directly use NOMA. The rest of this article is organized as follows. In Section \ref{sec:Waveform},  novel modulation candidates for OMA in 5G networks are compared. In Section \ref{sec:MA}, various NOMA schemes are discussed. Section \ref{sec:conclusion} concludes the article.

\section{Novel Modulation for OMA} \label{sec:Waveform}

In this section, we will discuss new modulation techniques for 5G networks. Since OFDM is widely used in current wireless systems and standards, many potential modulation schemes for 5G networks are delivered from  OFDM for backward compatibility reasons. Therefore, we will first introduce traditional OFDM.

\subsection{Traditional OFDM}	\label{sec:traditionalofdm}
Denote $d_k$, for $k=0,1,\cdots,N-1$, to be the transmit complex symbols. Then the baseband OFDM signal can be expressed as
\begin{equation}
s(t) = \sum_{k=0}^{N-1} d_{k} e^{j2\pi f_k t} \label{eq:ofdmsig}
\end{equation}
for $0 \leq t \leq T_s$, where $f_k=k \Delta f$, $\Delta f$ is the subcarrier bandwidth and $T_s$ is the symbol duration. To ensure that transmit symbols can be recovered without distortion, $\Delta f\cdot T_s=1$, which is also called the orthogonal condition. It can be easily shown that
\begin{equation}
d_{k} = \frac{1}{T_s} \int_0^{T_s} s(t) e^{-j2\pi f_k t} \, dt \label{eq:ofdmsym}
\end{equation}
if the orthogonal condition holds.

Denote $s(n \Delta t)$ to be the sampled version of $s(t)$, where $\Delta t=\frac{Ts}{N}$. It can be easily seen \cite{book}  that $\{s(0),s(\Delta t),\cdots,s((N-1)\Delta t)\}$ is the inverse discrete Fourier transform (IDFT)  of $\{d_0,d_1,\cdots,d_{N-1}\}$, which can be implemented by fast Fourier transform (FFT) and significantly simplifies OFDM modulation and demodulation.

To address the delay spread of wireless channels, a CP is usually used in OFDM. If the length of the CP is larger than the delay span (the duration between the first  and the last taps/paths of a channel), then the demodulated OFDM  signal can be expressed as
\begin{equation}
\hat{d}_{k} = H_k d_k + n_k, \label{eq:ofdmdem}
\end{equation}
where $H_k$ is the frequency response of the wireless channel at $f_k=k\Delta f$ and $n_k$ is the impact of additive channel noise. Therefore, the channel distortion becomes a multiplication of channel frequency response in OFDM systems while it is convolution in single-carrier systems, which makes the detection of OFDM signal much easier.

From the above discussion, OFDM can effectively deal with the delay spread of broadband wireless channels and FFT can be used to significantly simplify its complexity, therefore it has been widely used in the current wireless communication systems and standards.  However, as we can see from (\ref{eq:ofdmsig}), the OFDM signal is time-limited. Therefore, its OOB leakage is pretty high, especially when users are asynchronized as typical of 5G networks. To address this issue, a guard band is usually inserted between the signals of two adjacent users in the frequency domain in addition to a CP or a guard interval in the time domain, which reduces the SE of OFDM.  This is even more severe  for the users using a narrow frequency band.

5G networks have to support not only a massive number of users but also dramatically different types of users that have different demands. Traditional OFDM can no longer satisfy these requirements, and therefore novel modulation techniques  with much lower OOB leakage are required. The new modulation techniques for 5G networks currently need to consider backward compatibility with traditional OFDM systems but should also have the following key features to address the new challenges.
\begin{enumerate}
	\item {High SE}: New modulation techniques should be able to mitigate OOB leakage among adjacent users so that the system SE can be improved significantly  by reducing the guard band/time resources.
	\item {Loose synchronization requirements}: Massive number of users are expected to be supported, especially for the Internet of things (IoT), which makes synchronization difficult. Therefore, new modulation techniques are expected to accept asynchronous scenarios.
	\item {Flexibility}: The modulation parameters (e.g., subcarrier width and symbol period) for each user should be configured independently and flexibly to support users with different data rate requirements.
\end{enumerate}

The modulation techniques for OMA mainly include pulse shaping, subband  filtering, precoding design,  guard interval (GI) shortening, and modulation in the delay-Doppler domain. In this section, we  introduce those promising modulation techniques subsequently.

\subsection{Modulations based on Pulse Shaping}	\label{sec:pulseshaping}
Pulse shaping, which is also regarded as subcarrier-based filtering, can effectively reduce OOB leakage.
%
According to the Heisenberg-Gabor uncertainty principle \cite{uncertainty},  the time and frequency widths of the pulses cannot be reduced at the same time. Therefore, the waveforms based on pulse shaping is usually non-orthogonal in both time and frequency domains to maintain high SE.
Compared with traditional OFDM, the transceiver structure supporting pulse shaped modulation is more complex.
Here, we introduce two typical modulations based on pulse shaping, i.e., filter bank multicarrier (FBMC) \cite{Farhang2011,PPN} and generalized frequency division multiplexing (GFDM) \cite{GFDM2,GFDM}.

\subsubsection{FBMC}	\label{sec:FBMC}

\begin{figure}[!t]
	\centering
	\includegraphics[width=6.85in]{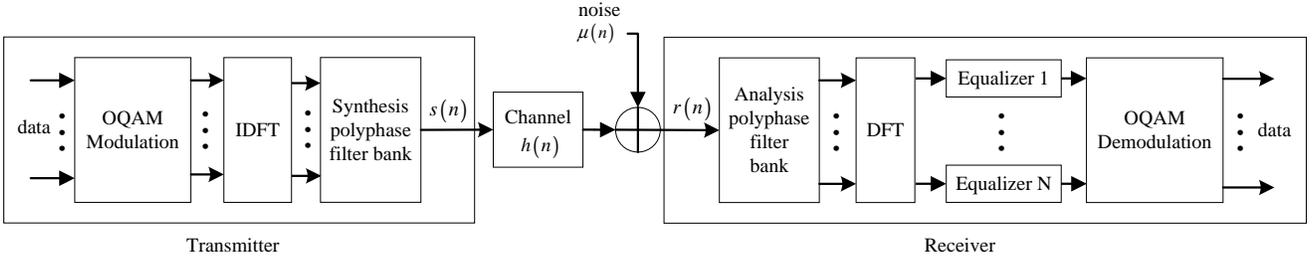}
	\caption{\small{Filter bank multicarrier (FBMC).}}
	\label{fig:fbmc}
\end{figure}


As shown in Fig. \ref{fig:fbmc}, FBMC \cite{Farhang2011,PPN} consists of IDFT and DFT, synthesis and analysis polyphase filter banks.
The prototype filter in FBMC performs the pulse shaping. There are two types of typical pulses: the pulse based on the isotropic orthogonal transform algorithm (IOTA) \cite{IOTA} and the pulse adopted in the PHYDYAS project \cite{PHYDYAS}. The length of the pulse in the time domain is determined by the required performance and is usually several times the length of the symbol period. The bandwidth of the pulse, which is different from the pulse in the traditional OFDM that has a long tail, is limited within a few subbands. To achieve the best SE, offset quadrature amplitude modulation (OQAM) is usually applied to make FBMC real-domain orthogonal in time and frequency domains~\cite{Farhang2011}. Therefore, the transmit signal over $M/2$ consecutive block periods\footnote{Block period corresponds to OFDM symbol period in the traditional OFDM systems.} can be expressed as
\begin{equation}
s(n) = \sum_{k=0}^{K-1} \sum_{m=0}^{M-1} d_{k,m} \theta_{k,m} g \left( n-mK/2 \right) e^{\frac{j2\pi kn}{K}}, \label{eq:fbmcsig}
\end{equation}
where
$K$ and $M$ are the numbers of subcarriers and symbols, respectively, $d_{k,m}$ is the transmit symbol at subcarrier $k$ and symbol $m$, and $g(n)$ is the prototype filter coefficient at the $n$-th time-domain sample. It is worth noting that the transmit symbols here refer to the pulse amplitude modulation (PAM) symbols that are derived from the staggering of quadrature amplitude modulation (QAM) symbols. Thus the interval between two adjacent blocks  is only half of the block period due to the offset in OQAM. The parameter, $\theta_{k,m}$ in (\ref{eq:fbmcsig}), is defined as
\begin{equation} \label{eq:theta}
\theta_{k,m} = \begin{cases}
\pm 1, & \text{if } m+k \text{ is even,} \\
\pm j, & \text{if } m+k \text{ is odd,}
\end{cases}
\end{equation}
which is used to form the OQAM structure.  With a properly designed prototype filter such as IOTA and the OQAM structure, the interference from the nearby overlapped symbols caused by a matched filter (MF) receiver becomes pure imaginary, which can be easily cancelled.

\subsubsection{GFDM}	\label{sec:GFDM}
\begin{figure}[!t]
	\centering
	\includegraphics[width=6.2in]{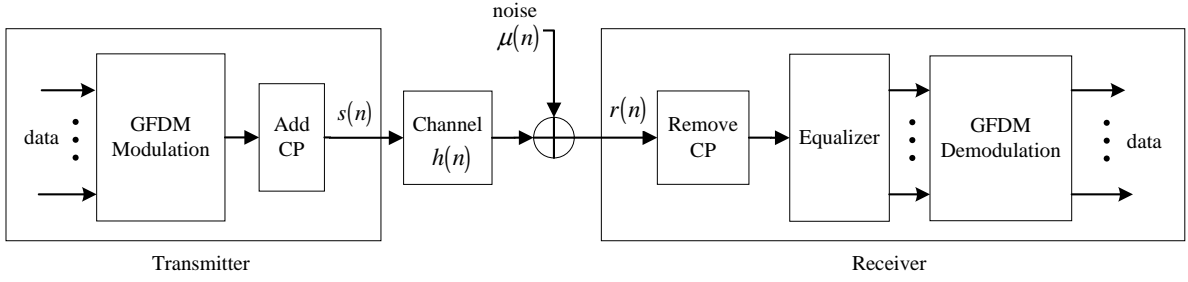}
	\caption{\small{Generalized frequency division multiplexing (GFDM).}}
	\label{fig:gfdm}
\end{figure}

Fig. \ref{fig:gfdm} demonstrates the block diagram of GFDM. OFDM and single-carrier frequency division multiplexing (SC-FDM) can be regarded as two special cases of GFDM \cite{GFDM}.
The unique feature of GFDM is to use circular shifted filters, rather than linear filters that are used in FBMC, to perform pulse shaping. By carefully choosing the circular filter, the out-of-block leakage can be reduced even if the orthogonality is completely given up.
We can flexibly adjust $M$ frequency samples and $K$ time samples for a GFDM block according to the application environment.
The transmit signal for each GFDM block can be expressed as
\begin{equation}
s(n) = \sum_{k=0}^{K-1} \sum_{m=0}^{M-1} d_{k,m} g_{k,m}(n),
 \label{eq:gfdmsig}
\end{equation}
for $0\leq n \leq KM-1$, where
$d_{k,m}$ is the transmit symbol on subcarrier $k$ at subsymbol $m$ and $g_{k,m}(n)$ is the circular time and frequency shifted version of the prototype pulse shaping filter.
In (\ref{eq:gfdmsig}),
\begin{equation}
g_{k,m}(n) = g \left( (n-mK)_{KM} \right) e^{\frac{j2\pi kn}{K}},
\end{equation}
where $(.)_{KM}$ denotes the $KM$ modulo operation and $g(n)$ is the prototype pulse shaping filter.
Similar to the traditional OFDM, the modulation process and demodulation process can be expressed by matrix operations. The IDFT and DFT matrices in the traditional OFDM are substituted by some specific matrices corresponding to the modulation and demodulation for GFDM. But, the transceiver structure of GFDM is significantly different from the traditional OFDM.

Besides FBMC and GFDM, other modulations based on pulse shaping, such as pulse-shaped OFDM \cite{pofdm} and QAM-FBMC~\cite{QAM-FBMC}, have also been  proposed for 5G networks.
Generally, modulations based on pulse shaping try to restrict transmit signals within a narrow
bandwidth  and thus mitigate the OOB leakage so that they can work in asynchronous scenarios with a narrow guard band.
FBMC also uses OQAM to achieve real-domain orthogonality, which saves the cost of the GI and interference cancellation. In addition, the circular shifted filters in GFDM avoid the long tail of the linear filters in the time domain, which makes GFDM fit for sporadic transmission. Furthermore, GFDM is easily compatible to MIMO technologies \cite{GFDM}.






\subsection{Modulations based on Subband Filtering}	\label{sec:subbandfiltering}
Subband filtering is another technique to reduce the OOB leakage.
Universal filtered multicarrier (UFMC)~\cite{UFMC,UFMCvsFBMC2} and filtered OFDM (f-OFDM)~\cite{fofdm,fofdm2} are two typical modulations based on subband filtering, which will be introduced next.

\subsubsection{UFMC}	\label{sec:UFMC}

\begin{figure}[!t]
	\centering
	\includegraphics[width=6.7in]{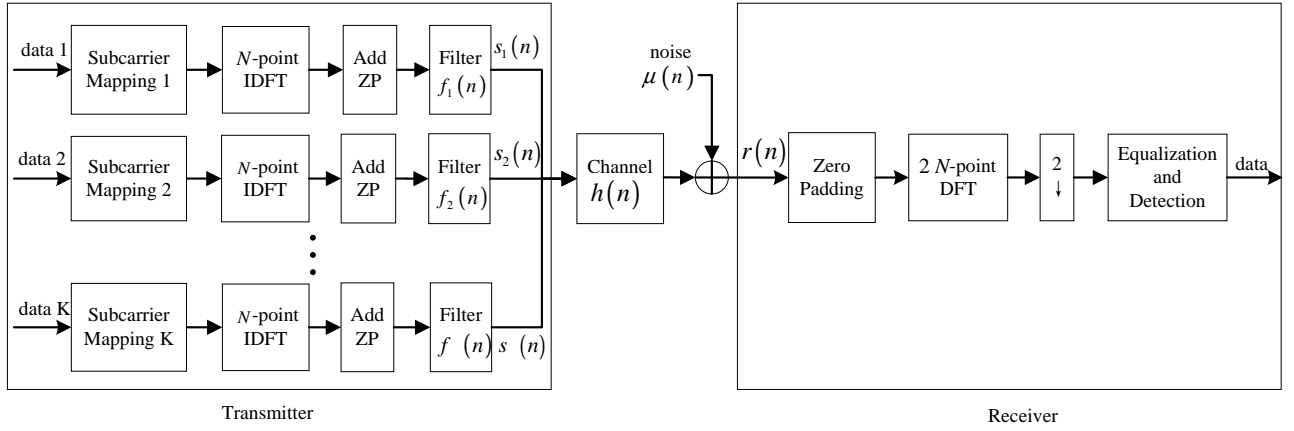}
	\caption{\small{Universal filtered multicarrier (UFMC).}}
	\label{fig:ufmc}
\end{figure}

Fig. \ref{fig:ufmc} shows the transmitter and the receiver structures of UFMC \cite{UFMC,UFMCvsFBMC2}.
In UFMC, the subbands are with equal size, and each filter is a shifted version of the same prototype filter. OFDM is applied  within a subband for this modulation as shown in the figure.
Since the bandwidth of the filter in UFMC is much wider than that of the modulations based on the pulse shaping, the length in time domain is much shorter. Therefore, interference caused by the tail of the filter can be easily eliminated by  adopting a zero-padding (ZP) prefix with a reasonable length. Assuming that $N$ subcarriers are divided into $K$ subbands, each with  $L=N/K$ consecutive subcarriers, the transmit signal in UFMC can be expressed as
\begin{equation}
s(n) = \sum_{k=0}^{K-1} s_{k}(n) * f_{k}(n), \label{eq:ufmcsig}
\end{equation}
where $f_{k}(n)$ is the filter coefficient of subband $k$, and $s_{k}(n)$ is the OFDM modulated signal over subband $k$ that can be expressed as
\begin{equation}
\begin{split}
s_{k}(n) &= \sum_{m=0}^{M-1} s_{k,m}(n-m(N+N_g)) \\
&= \begin{cases}
s_{k,m}(n-m(N+N_g)), &  m(N+N_g) \leq n \leq m(N+N_g)+N-1 \\
0, & \text{otherwise}
\end{cases}
\label{eq:ufmcsubsig}
\end{split}
\end{equation}
with $N_g$ denoting the length of the ZP, $M$ denoting the number of symbol blocks and $s_{k,m}(n)$ denoting the signal at subcarrier $k$ and symbol $m$.
In (\ref{eq:ufmcsubsig}), $s_{k,m}(n)$ can be expressed as
\begin{equation}
s_{k,m}(n) = \sum_{l=(k-1)L}^{kL-1} d_{l,m} e^{j\frac{2\pi ln}{N}} , \quad 0\leq n \leq N-1, \label{eq:ufmcsubsubsig}
\end{equation}
where $d_{l,m}$ denotes the $l$-th transmit symbol at the $m$-th symbol block.

At the receiver, 
the signal at each symbol interval is with the length of $N+N_g$ and is zero-padded to have a length of $2N$ so that a $2N$-point FFT can be performed.
Please note that only the even subcarriers are considered for signal detection after the $2N$-point FFT.



\subsubsection{f-OFDM}	\label{sec:F-OFDM}

f-OFDM has a similar transmitter structure as UFMC \cite{fofdm,fofdm2}. The main difference is that f-OFDM employs a CP and usually allows
residual inter-symbol interference (ISI) \cite{fofdm}.
Therefore, at the receiver, the MF is applied instead of the ZP and decimation.
Besides, downsampling can be applied before the DFT operation, which can reduce complexity significantly since
the CP can mitigate most of interference caused by the tail of the filter; the residual interference is with much lower power and can be treated as noise \cite{fofdm2}. Thus, the filter in f-OFDM can be longer than that in UFMC and has better attenuation outside the band. With the aid of effective channel coding, the performance degradation caused by residual interference in f-OFDM can be negligible.
Another difference from UFMC is that the subcarrier spacing and the CP length do not have to be the same for different users in f-OFDM.

The most widely used filter in f-OFDM is the soft-truncated sinc filter \cite{fofdm},
which can be easily used in various applications with different parameters. Therefore, f-OFDM is very flexible in the frequency multiplexing.


Besides UFMC and f-OFDM, other modulations based on subband filtering have also been proposed. For example, resource block f-OFDM (RB-f-OFDM) \cite{rbfofdm} utilizes filters based on resource block instead of the whole band of users in f-OFDM. In general, modulations based on subband filtering can effectively reduce OOB leakage and achieve better performance in comparison with the traditional OFDM.

\subsection{Other Modulation Techniques}	\label{sec:othertech}
Apart from pulse shaping and subband filtering, there are also some other techniques to suppress the OOB leakage and meet the requirements of 5G networks. In the following, we mainly introduce three other modulations, including guard interval discrete Fourier transform spread OFDM (GI DFT-s-OFDM) \cite{GI-DFT-s-OFDM}, spectrally-precoded OFDM (SP-OFDM) \cite{spofdm}, and orthogonal time frequency and space (OTFS) \cite{OTFS}.

\subsubsection{GI DFT-s-OFDM}	\label{sec:GI-DFT-s-OFDM}
In GI DFT-s-OFDM \cite{GI-DFT-s-OFDM},
the known sequence is used as the GI instead of a CP.
Several types of the known sequences, such as the zero sequence \cite{ZT-DFT-s-OFDM} and a well-designed unique word \cite{UW-DFT-s-OFDM}, can be used.
By a fixed known sequence with constant amplitude in GI DFT-s-OFDM,
the peak-to-average power ratio (PAPR) of the modulated signal can be reduced. Moreover, the known sequence can also be utilized to estimate the parameters, such as the carrier frequency offset (CFO) in the synchronization process.
By utilizing a proper sequence as the GI, the discontinuity between the adjacent time blocks in the traditional OFDM/DFT-s-OFDM can be avoided. As a result, the OOB leakage is reduced.

For GI DFT-s-OFDM, the overall length of the GI and useful signal for different users is same.
Thus, the DFT windows for different users at the receiver can still be aligned even if the lengths of the GIs are different. Therefore, the mutual interference due to asynchronization of users can be mitigated~\cite{GI-DFT-s-OFDM}.

\subsubsection{SP-OFDM}	\label{sec:spofdm}
\begin{figure}[!t]
	\centering
	\includegraphics[width=6in]{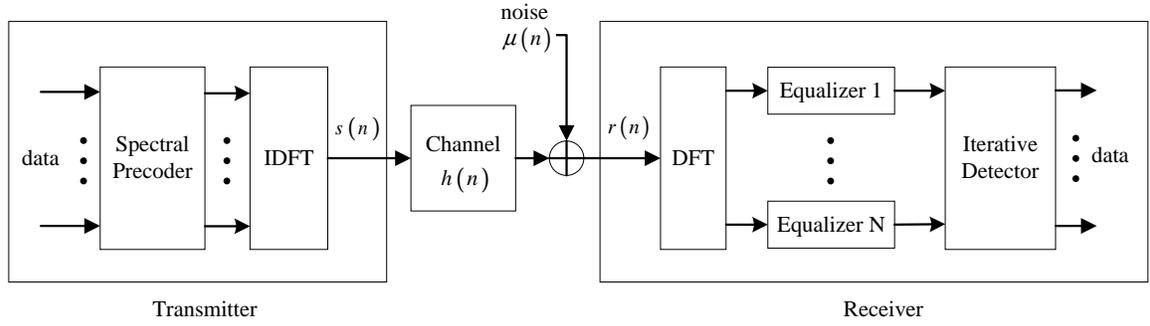}
	\caption{\small{Spectrally-precoded OFDM (SP-OFDM).}}
	\label{fig:spofdm}
\end{figure}

Fig. \ref{fig:spofdm} shows the diagram of SP-OFDM \cite{spofdm}. From the figure, it consists IDFT and DFT, spectral precoder, and iterative detector.
Generally, the  data symbols mapped on subcarriers are precoded by a rank-deficient matrix in order to project the signal into a properly selected lower dimensional subspace so that the precoded signal can be high-order continuous, and results in much lower leakage compared with the traditional OFDM \cite{ncontiofdm,ncontiofdm2}. Even if precoded by a rank-deficient matrix can reduce the capacity of the channel,  the OOB leakage of the OFDM signals can be significantly suppressed at the cost of only few reduced dimensions.


Compared to the modulations based on filtering, SP-OFDM has the following three advantages:
\begin{itemize}
\item The ISI caused by the tail of the filters can be removed without filtering. Therefore, the CP applied to combat the multipath of the wireless channels can be shorter, and SE is improved.
\item When fragmented bands are used, SP-OFDM can easily notch specific well chosen frequencies without requiring multiple narrow subband filters \cite{notchofdm}.
\item Furthermore, precoding and filtering can be combined to further improve the performance.
\end{itemize}

\subsubsection{OTFS}	\label{sec:OTFS}

The structure of OTFS is similar to SP-OFDM, as can be seen in Fig. \ref{fig:spofdm}. The main difference is that the spectral precoder and the iterative detector  are substituted by the two-dimensional (2D) symplectic Fourier transform and the corresponding inverse transform modules. OTFS maps the symbols in the delay-Doppler domain \cite{OTFS}. Through a 2D symplectic Fourier transform, the corresponding data in the time-frequency domain can be calculated. Then, the calculated data can be transmitted via a time-frequency-domain modulation method as in OFDM. Since the 2D symplectic Fourier transform is relatively independent of the time-frequency-domain modulation method, pulse shaping and subband filtering can also be applied together to further reduce the leakage in OTFS.

When a mobile is with a high speed, the channel experiences fast fading. Channel parameters need to be estimated and tracked very often therefore, which significantly increases resource costs . Moreover, most of the modulations are designed assuming that channels are constant within a symbol block. With a high mobility speeds, extra interference is introduced, which degrades the performance. However, in the delay-Doppler domain, the high Doppler channel can be expressed in a stable model, which saves the cost of tracking the time-varying fading and improves performance therefore.

OTFS can be also applied to estimate channel state information (CSI) of different antennas in MIMO systems \cite{OTFS}.  Generally, the delay and Doppler dispersions are  still relatively small compared to the system scale. In this case, the channel can be expressed in a compact and stable form in the delay-Doppler domain. As a result, the spread of the pilots caused by the channel are local, which enables  to estimate the CSI of different antennas in MIMO systems by different pilots within a small area of the delay-Doppler plane. 


 In addition, a number of modulations based on other techniques have been also proposed, such as windowed OFDM (W-OFDM) \cite{wofdm}, which utilizes windowing to deal with the discontinuity between adjacent OFDM symbols.




\subsection{Performance Comparison of Different Modulations}	\label{sec:wavecomp}
We compare the power spectral density (PSD) and bit-error rate (BER) of different modulations.

\subsubsection{PSD}	\label{sec:psd}
\begin{figure}[!t]
	\centering
	\includegraphics[width=3.5in, height=2.5in]{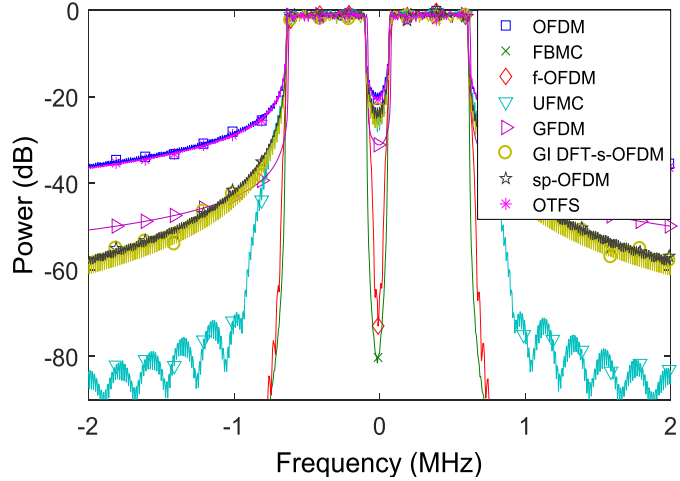}
	\caption{\small{PSDs of different modulations, where the bandwidth is 15.36 MHz, the DFT size is 1024, each of the two users occupies 36 subcarriers, and the guard band contains 12 subcarriers. }}
	\label{fig:psd}
\end{figure}

Suppressing the OOB leakage is a key purpose for most of the modulation candidates for 5G networks.
The PSDs of the some modulations are shown in Fig. \ref{fig:psd}. From the figure, all modulations achieve much lower leakage compared to the traditional OFDM. Among them, UFMC applies subband filtering and also has  low leakage, and
FBMC and f-OFDM have the lowest leakage.
GFDM, GI DFT-s-OFDM, and SP-OFDM, although do not reduce the leakage as much as FBMC and f-OFDM, can still achieve much better performance than the traditional OFDM.

\subsubsection{BER}	\label{sec:ber}
In order to reduce the OOB leakage, many modulations utilize techniques, such as pulse shaping and subband filtering, which may introduce ISI and ICI. Hence, the BER performance of different modulations is compared here.

\begin{figure}[!t]
	\centering
	\subfloat[$f_d=0$ Hz]{\includegraphics[width=3.2in,height=2.3in]{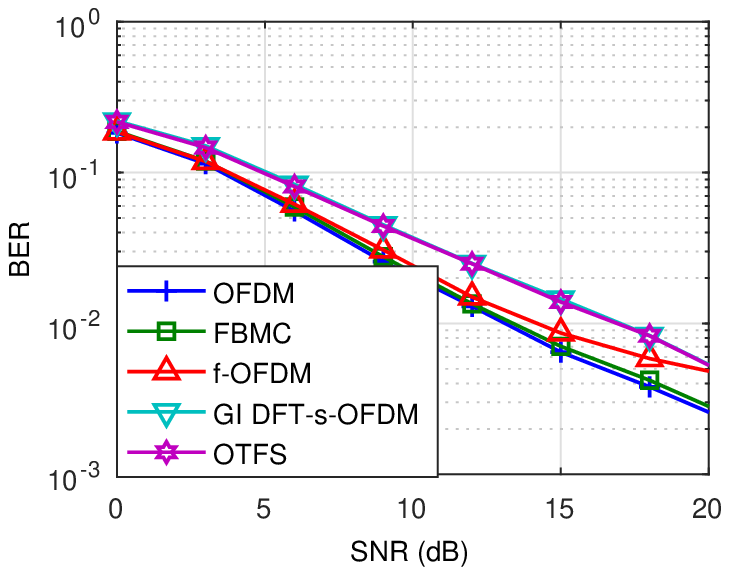} \label{fig:bera}}
	\hfill
	\subfloat[$f_d=300$ Hz]{\includegraphics[width=3.2in,height=2.3in]{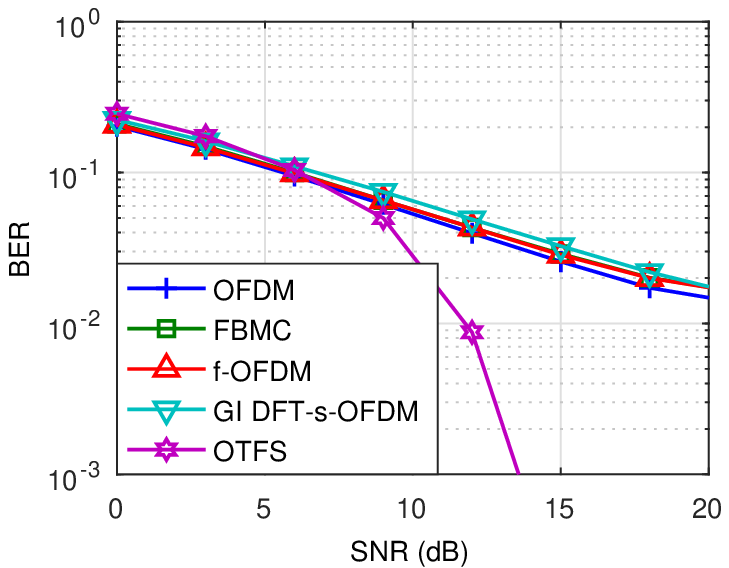} \label{fig:berb}}
	\caption{\small{BER versus SNR for different modulations, where the bandwidth is 15.36 MHz, the DFT size is 1024, each of the two users occupies 36 subcarriers, and the guard band contains 12 subcarriers for ITU Vehicular A channel model.}}
	\label{fig:ber}
\end{figure}

 Fig. \ref{fig:ber} shows the BER performance versus signal-to-noise ratio (SNR) when the Doppler spread $f_d=0$ and $300$ Hz.
 From Fig.~\ref{fig:ber}~(a), the traditional OFDM has the best performance when the Doppler spread is zero ($f_d=0$) since the ISI caused by the multipath has been completely canceled by the CP.
 Since the bandwidth of each subcarrier is small enough to make the corresponding channel approximately flat, the ISI introduced by pulse shaping in FBMC is nearly pure imaginary. Therefore, FBMC is approximately orthogonal in the real domain and achieves good BER performance.
 The performance of UFMC, GFDM, and SP-OFDM is similar to that of FBMC, which is degraded slightly due to noise enhancement
and low-projection precoding. However, f-OFDM introduces extra ISI that cannot be completely canceled, and as a result, it has slightly worse performance, especially in the high SNR region.  GI DFT-s-OFDM and OTFS, which are different from the modulation schemes that directly map the symbols on subcarriers, apply spreading before mapping  so that their performance does not approach that of OFDM.
 Since the fast-fading channel  is difficult to be estimated and tracked accurately, the performance of the most modulation schemes degrades significantly as we can see from Fig. \ref{fig:ber} (b). While OTFS can still achieve good performance due to its specific channel estimation method. Moreover, its performance in the high-mobility scenario is even  better than that in the zero Doppler shift scenario because of Doppler diversity.



\subsection{Open Issues}	\label{sec:futureworkwav}
In this section, modulation techniques for 5G networks will be be discussed.
These techniques can be used with OMA to effectively deal with the OOB leakage in  5G networks. However, there are still many open issues in the area.

A potential application of f-OFDM is IoT. In this scenario, the subbands are narrow and therefore, interference caused by a short CP can  significantly degrade the performance and should be considered in the detection. To improve the detection performance, additional processing, such as filtering or successive interference cancellation (SIC), is needed. Residual ISI cancellation (RISIC)~\cite{RISIC} could be helpful.

The existing designs for subband filtering, such as Dolph-Chebyshev filter in UFMC and soft-truncated filter in f-OFDM, are with fixed length. However, different users and different application scenarios will have different requirements on the leakage levels, filter lengths, etc. According to the Heisenberg-Gabor uncertainty principle~\cite{uncertainty}, the time and frequency dispersions are dual variables that cannot be reduced at the same time. Therefore, how to balance the time and frequency dispersions and  design an efficient prototype filter according to  application scenarios is interesting.

Similar to the traditional OFDM, multi-carrier based new modulation candidates, such as FBMC and UFMC also have a large PAPR. In order to improve the efficiency of the power amplifier, the PAPR should be reduced. The traditional PAPR reduction methods~\cite{PAPRred} applied in the traditional OFDM usually introduce distortions that degrade the performance. Therefore, how to properly extend the PAPR reduction methods in the traditional OFDM to the new modulations is an interesting and meaningful issue.



\section{NOMA} \label{sec:MA}
In order to support higher throughput and massive and heterogeneous connectivity for 5G networks, we can adopt novel modulations discussed in Section II for OMA, or directly use NOMA with effective interference mitigation and signal detection methods.  The key features of NOMA can be summarized as follows:
\begin{enumerate}
\item {Improved SE}: NOMA exhibits a high SE, which is attributed to the fact that it allows each resource block  (e.g., time/frequency/code) to be exploited by multiple users.
\item {Ultra high connectivity}: With the capability to support multiple users within one resource block, NOMA can potentially   support massive connectivity for billions of smart devices. This feature is quite essential for IoT scenarios with users that only require very low data rates but with massive number of users.
\item {Relaxed channel feedback}: In NOMA, perfect uplink CSI is not required at the base station (BS). Instead, only the received signal strength needs to be included in the channel feedback.
\item {Low transmission latency}: In the uplink of NOMA,  there is no need to schedule requests from  users to the BS, which is normally required in OMA schemes. As a result, a grant-free uplink transmission can be established in NOMA, which  reduces the transmission latency drastically.
\end{enumerate}

Existing NOMA schemes can be classified into three categories: power-domain NOMA, code-domain  NOMA, and NOMA multiplexing in multiple domains. We will introduce them subsequently
with emphasis on power-domain NOMA.

\subsection{Power-Domain NOMA}	\label{sec:NOMA}
Power-domain NOMA is considered as a promising MA scheme for 5G networks \cite{higuchi2013non}. Specifically,  a downlink version of NOMA, named multiuser superposition transmission (MUST), has been proposed for the 3GPP long-term evolution advanced (3GPP-LTE-A) networks~\cite{scheme1-1}. It has been shown that system capacity and user experiences can be improved by NOMA. More recently, a new work item (WI) outlining downlink multiuser superposition transmission for LTE has  been approved by 3GPP LTE Release 14~\cite{scheme2}, which aims to identify the necessary techniques to enable LTE to support  the downlink intra-cell multiuser superposition transmission. Here, we will expand upon the basic principles of various power-domain NOMA related techniques, including multiple antenna based NOMA, power allocation in NOMA, and cooperative NOMA.

\subsubsection{Basic Power-Domain NOMA}
\begin{figure*}[!t]
	\centering
	\includegraphics[width=5in]{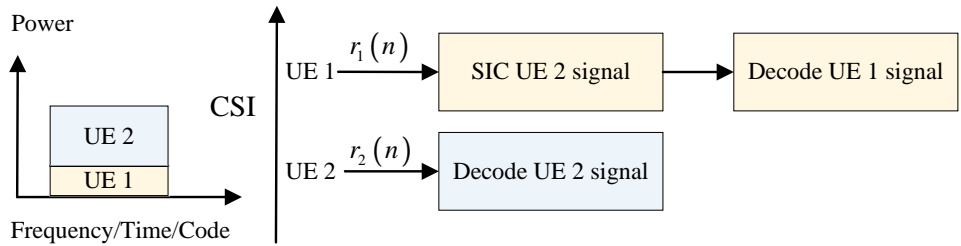}
	\caption{\small{Downlink power-domain non-orthogonal multiple access (NOMA).}}
	\label{fig:noma}
\end{figure*}

Power-domain NOMA, as illustrated in Fig.~\ref{fig:noma} for the two user case, deviates from  conventional OMA that uses TDMA/FDMA/CDMA/OFDMA allocating orthogonal resource blocks for different users to avoid the multiple access interference (MAI). Instead power-domain NOMA can support multiple users within the same resource block by distinguishing them with different power levels. As a result, NOMA is able to support more connectivity and provide higher throughput with limited resources.

The downlink transmission of NOMA for the two user case is shown in Fig.~\ref{fig:noma} where the users are served at the same time/frequency/code resource block with a total power constraint. Specifically, the BS sends a superimposed signal containing the two  signals for the two users. This differs from conventional power allocation strategies, such as water filling,as NOMA allocates less power for the users with better downlink CSI, to guarantee overall fairness and to utilize diversity in the time/frequency/code domains.  SIC is used for signal detection at the receiver. The user with more transmit power, that is, the one with smaller downlink channel gain, is first to be decoded while treating the other user's signal as noise. Once the signal corresponding to the user with the larger transmit power is detected and decoded, its signal component will be subtracted from the received signal to facilitate the detection of subsequent users. It should be noted that the first detected user is with the largest inter-user interference and also the detection error in the first user will pass to the other user, which is why we have to allocate sufficient power to the first user to be detected. The extension of NOMA from two to multiple user cases is straightforward.

For the uplink transmission of NOMA, the transmit power is limited by each individual user. Different from the downlink, the transmit powers of the users using the same resource block are carefully controlled so that the received signal components at the BS corresponding to the users with the better CSI, have more powers. At the receiver (the BS), the user with the best CSI is decoded first. After that, the corresponding component is removed from the received signal. The SIC receiver works in a descending order of the CSI, which is the opposite to the downlink case.

\begin{figure}[!t]
	\begin{center}
	\includegraphics[width=3.8in, height=2.7in]{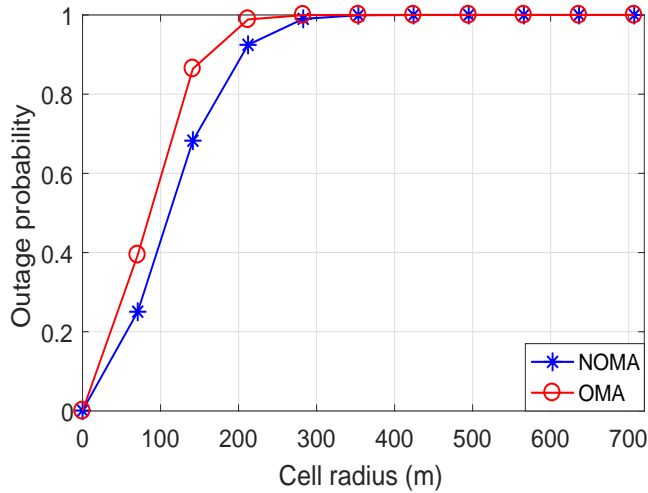}
	\end{center}
	\caption{\small{Outage probability of NOMA and OMA with different cell radius, two users are supported by NOMA scheme where the far user is located at the edge of the cell, and the one user supported by OMA scheme is randomly located in the cell. The path loss exponent is 2. The transmit power signal-to-noise (SNR) is 40 dB. Power allocation coefficients for two users in NOMA are $4/5$ and $1/5$, respectively.}}
	\label{fig:NOMA_OMA}
\end{figure}

Fig.~\ref{fig:NOMA_OMA} compares  NOMA and OMA  where two users are served by the same BS if NOMA is adopted. From the figure, the NOMA scheme achieves a lower outage probability. However, by adopting NOMA, a more complex transmitter and receiver are required to mitigate the interference. Furthermore, power-domain NOMA usually works well when only two or a few users share the same resource block. As the number of users multiplexing in power domain increases, the MAI becomes severe and the performance of NOMA degrades.

\subsubsection{Multiple Antenna based NOMA}
Multiple antenna techniques can provide an additional degree of freedom on the spatial domain, and bring further performance improvements to NOMA. Recently, multiple antenna based NOMA has attracted lots of attention~\cite{higuchi2013non,Jinho:2015,higuchi2015non,Chen:TSP:2016,ding2015mimo,Zhiguo_general:2015,Zhiguo2016IoT,Sunqi2015,Ding2016SPL,Yuanwei2016NOMA,Zhijin:ICC:2016,Yuanwei:twc:2017}. Different from single-input-single-output (SISO) based NOMA, where the channels are normally represented by scalars, one of the research challenges in multiple antenna based NOMA comes from user ordering; as the channels are generally in form of vectors or matrices. Currently, the possible designs of multiple antenna based NOMA fall into two categories where one or multiple users are served by a single beamforming vector.

By allocating different users with different beams in the same resource block, the quality of service (QoS) of each user can be guaranteed in multiple antenna based NOMA systems forcing the beams to satisfy a predefined order. This type of multiple antenna based NOMA scheme has been first proposed by Sun~\emph{et al.} in~\cite{Sunqi2015}  to investigate power optimization to maximize the ergodic capacity. This proposed multiple antenna based NOMA scheme has proved to be able to achieve significant performance improvement compared with conventional OMA schemes.

A cluster of users can share the same beam. The spatial channels of different users within the same cluster are considered to be highly correlated. Therefore, beams for different clusters should be carefully designed to guarantee that the channels for different clusters are orthogonal to each other in order to suppress the inter-cluster interference. For multiple-input-single-output (MISO) based NOMA, a two-stage multicast beamforming scheme has been proposed by Choi in~\cite{Jinho:2015}, where ZF beamforming has been employed  to mitigate interference from adjacent clusters first and then the optimal beamforming vectors have been designed to minimize the total transmit power within each cluster. For MIMO based NOMA, a scheme to simultaneously apply open-loop random beamforming and intra-beam SIC, has been proposed by Higuchi and  Kishiyama in~\cite{higuchi2013non}. However, here the system performance is considerably degraded as the random beamforming can bring uncertainties at the user side. More recently, a precoding and detection framework with fixed power allocation has been proposed by Ding~\emph{et al.}~\cite{ding2015mimo} to solve these problems caused by random beamforming, and demonstrated that MIMO based NOMA can achieve better outage performance than MIMO based OMA even for users who experience strong co-channel interference.

A comprehensive summary for the state-of-the-art work on multiple antenna based NOMA is given in Table~\ref{table:MIMO NOMA}, where ``BF'', ``OP'', ``SU'' and ``MU'' are used to represent  beamforming, outage probability, two-user and multi-users cases, respectively.

\begin{table}[!t]
\caption{\small{Multiple antenna based NOMA}}
\begin{center}
\centering
\begin{tabular}{|l|l|l|l|l|}
\hline
\centering
\textbf{Scenarios}  & \textbf{Techniques}& \textbf{Metrics} & \textbf{Characteristics} & \textbf{Ref.}\\
\hline
\centering
MU-MISO  & Two stage BF & Transmit power/rate & Optimal power allocation  & \cite{Jinho:2015} \\
\hline
\centering
 MU-MISO  & Restrictive BF & Throughput & Less constraints at the BS & \cite{higuchi2015non}\\
\hline
\centering
 MU-MISO &  ZF-BF + DPC & Power consumption \& OP & Precoding with low complexity \& user pairing &   \cite{Chen:TSP:2016} \\
\hline
\centering
 MU-MIMO &  Random BF & Sum rates & No perfect CSI required& \cite{higuchi2013non} \\
\hline
\centering
 MU-MIMO &  ZF-BF & OP+rate gain & No CSI available at the transmitter  & \cite{ding2015mimo} \\
\hline
\centering
 MU-MIMO  & Signal alignment & OP  & Larger diversity gain &   \cite{Zhiguo_general:2015}\\
\hline
\centering
SU-MIMO &  QR based deposition & OP & Channel differences & \cite{Zhiguo2016IoT} \\
\hline
\centering
SU-MIMO & Beamformer-based & Ergodic capacity & Optimal/sub-optimal power allocation & \cite{Sunqi2015} \\
\hline
\centering
MU-MIMO &  Two-stage precoding & OP+sum rate & One bit feedback &   \cite{Ding2016SPL} \\
\hline
\centering
 MU-MIMO & ZF-BF & Max-min rate & Fairness in MIMO-NOMA &   \cite{Yuanwei2016NOMA}\\
\hline
\end{tabular}
\end{center}
\label{table:MIMO NOMA}
\end{table}

\subsubsection{Power Allocation in NOMA}
NOMA is capable of supporting unequal transmission rates for users experiencing various channel conditions by assigning them different transmit powers. Therefore, the power allocation mechanism for different users is critical to power-domain NOMA. In the downlink of NOMA, the power allocation  is the opposite to the conventional schemes (e.g., water filling policy), and more powers are allocated to the users with poor CSI in order to ensure that every user obtains reasonable receive signal power and therefore guarantee fairness. The optimization problem is normally modelled to maximize the individual/sum rate while considering this fairness issue. As the power allocation is based on the order of CSI, the cases with perfect and imperfect CSI are different and should be investigated separately, as in~\cite{Timotheou:2015}. When perfect CSI is available, the optimization problem can be formulated to maximize the minimum achievable user rates. With average CSI rather than perfect CSI, the optimization problem can be formulated to minimize the maximum outage probability. A comprehensive summary of the state-of-the-art work on the power allocation in NOMA with different fairness strategies is given in Table~\ref{table:PA NOMA}. It is worth noting that the most critical challenge for power allocation in NOMA comes from the non-convex property of the ordered power constraints, which makes the optimization problem untraceable. Therefore, further research work on the optimal ordering design can be expected.

\begin{table}[!t]
\caption{\small{Power allocation in NOMA}}
\begin{center}
\centering
\begin{tabular}{|l|l|l|l|l|}
\hline
\centering
 \textbf{Power Allocation Strategy}   &\textbf{Fairness} & \textbf{Complexity}&\textbf{Ref.} \\
\hline
\centering
Allocating more power for weak users & Moderate & Low&\cite{NOMA,saito2013system,ding2014performance,yuanwei_JSAC_2015}\\
\hline
\centering
Give absolute priority to the fairness  & High& High& \cite{Timotheou:2015,Yuanwei2016NOMA,Cui2016NOMA} \\
\hline
\centering
Maximize the geometric mean of user rates  & Moderate& Low& \cite{Liu2015PIMRC,Liu2016ICC,Mei2016ICC,Otao2015fairness} \\
\hline
\centering
Give  users weighted priorities   & Low & Low& \cite{sun2016optimal}\\
\hline
\centering
Tradeoff between throughput and fairness & --- & --- &\cite{al2014uplink,lei2016power,WPT_NOMA_2015}\\
\hline
\end{tabular}
\end{center}
\label{table:PA NOMA}
\end{table}


\subsubsection{Cooperative NOMA}
In cellular networks, a cell-edge user usually experiences a weaker received signal power and lower data rates compared to those near the BS. Relaying and coordinated multipoint (CoMP) transmission (and reception) techniques have been widely employed to increase the transmission rates for cell-edge users~\cite{Lee:CoMP:2012}.  The scenario with users transmitting at different rates naturally matches the application scenarios typical of NOMA.

The basic idea of relay-assisted NOMA is to use the users with the better CSI as the decode-and-forward (DF)/amplify-and-forward (AF) as relays  to improve the transmission rates of the users with poor CSI. A cooperative NOMA model supporting $M$ users with $M$ time slots  has been proposed in~\cite{Ding2015cooperative}. In the first time slot, the traditional non-cooperative NOMA scheme is conducted. In the second time slot, the user with the best CSI acts as the DF relay for the user with the second best CSI. In the following time slots, the user with the $m$-th best CSI works as the relay for the user with the subsequent worse CSI to improve the transmission rates.

CoMP transmission, where multiple BSs support cell-edge users together, is capable of improving the performance of cell-edge users.  NOMA has been first applied into CoMP by Choi~\cite{Jinho2014Comp}, where two coordinated BSs use Alamouti code  to support a cell-edge user in a NOMA channel. Subsequently, the effectiveness of NOMA in CoMP systems has been further demonstrated by Tian~\emph{et~al.}~\cite{Tian2016CoMP} in comparison with the conventional joint-transmission NOMA. A  coordinated direct and relay transmission (CDRT) scheme has also been considered by Kim and Lee in~\cite{Kim2015CoMP}, where the BS communicates with a near user and a relay simultaneously, invoking NOMA in the first time slot while communicating with a far user with the aid of the relay in the following time slots. This NOMA based CDRT scheme solves the main challenge by using the inherent property of
NOMA that allows a receiver to obtain side information such as other user's signal for interference cancellation.

A comprehensive summary of the existing work on NOMA in cooperative communications is given in Table~\ref{table:cooperative NOMA}, where ``OP'' represents the outage probability.

\begin{table*}[!t]
\caption{\small{Cooperative NOMA}}
\begin{center}
\centering
\begin{tabular}{|l|l|l|l|}
\hline
\centering
  \textbf{Techniques}& \textbf{Metrics} & \textbf{Characteristics} &\textbf{Ref.} \\
\hline
\centering
AF & OP+ergodic rate & Multiple antenna relay &\cite{Men2015CL} \\
\hline
\centering
DF & OP+throughput & Self-powered relays &\cite{yuanwei_JSAC_2015} \\
\hline
\centering
DF & OP+ergodic rate & Maximum diversity gain for all users &\cite{Ding2015cooperative}   \\
\hline
\centering
DF & OP & Two stage relay selection schemes &\cite{ding2016relay} \\
\hline
\centering
 CoMP & Sum rate  & New coordinated superposition coding &\cite{Jinho2014Comp} \\
\hline
\centering
CoMP+DF & OP+ergodic rate & Each stream has same diversity order &\cite{Kim2015CoMP}\\
\hline
\centering
DF & Sum rate & Novel two-stage power allocation &\cite{duan2016use}\\
\hline
\end{tabular}
\end{center}
\label{table:cooperative NOMA}
\end{table*}

\subsubsection{Spectral and Energy Efficiency in NOMA}
Recall that when designing 5G  networks,  SE  and energy efficiency (EE) are two important performance metrics. NOMA provides a promising solution due to its extra  degree of freedom over the power domain, especially suited for IoT networks that require massive connectivity but with low power consumption at sensor nodes. More specifically, NOMA has been considered to be able to boost the SE of 5G networks~\cite{NOMA,Kim2015CoMP,Yuanwei:TVT:2016,Yuanwei:GC:2016}. NOMA with the consideration of EE has also been investigated in~\cite{Fang:2016}, where a sub-optimal resource allocation algorithm has been proposed to maximize the EE of the systems. Moreover,  the EE of a NOMA network, subject to a minimum required data rate for each user, can be maximized using the approach in~\cite{Zhang:EE:TVT:2016}. Similar to most wireless networks~\cite{Chen:SE:EE:2011}, SE and EE cannot be achieved simultaneously in NOMA networks. Therefore we expect to see more work on  SE and EE tradeoffs in NOMA in the future. A more comprehensive review on the power-domain NOMA can be found in~\cite{Dobre:2016}.

\subsection{Code-Domain NOMA}
Code-domain NOMA can support multiple transmissions within the same time-frequency resource block by assigning different codes to different users. It has certain spreading gain and shaping gain with the cost of extra signal bandwidth in comparison with  power-domain NOMA. Existing solutions to code-domain NOMA mainly include low-density spreading CDMA (LDS-CDMA)~\cite{Hoshyar:LDS-CDMA:2008}, low-density spreading OFDM (LDS-OFDM)~\cite{Mohammed:LDS-OFDM:2014}, and sparse code multiple access (SCMA)~\cite{SCMA}, which will be introduced in the following.

\subsubsection{LDS-CDMA}
LDS-CDMA~\cite{Hoshyar:LDS-CDMA:2008} is a novel type of CDMA. Its key feature is that a low-density signature, which has a similar form of the low-density parity-check (LDPC) matrix, is employed for the codebook construction. When the number of users is larger than that of samples per symbol period in the conventional CDMA,  MAI is inevitable and optimal multiuser detection is extremely complex. However, due to the sparse structure of the signature in LDS-CDMA, a low-complexity near-optimal multiuser detection scheme, based on a message passing algorithm (MPA), can be applied in the detection of LDS-CDMA, which significantly improves performance.

\subsubsection{LDS-OFDM}
LDS-OFDM~\cite{Mohammed:LDS-OFDM:2014} has similar properties to LDS-CDMA, except that the output of the signature is mapped into the subcarriers of OFDM rather than the time samples in CDMA. Therefore, a low-complexity MPA detector can be adopted. Compared to LDS-CDMA, LDS-OFDM utilizes multicarrier transmission, which makes it fit for wideband channels. Further, the strong compatibility with OFDM makes it flexible in resource allocation~\cite{Mohammed:LDS-OFDM:2014}.

\subsubsection{SCMA}	\label{sec:SCMA}
\begin{figure}[!t]
	\centering
\includegraphics[width=5.7in]{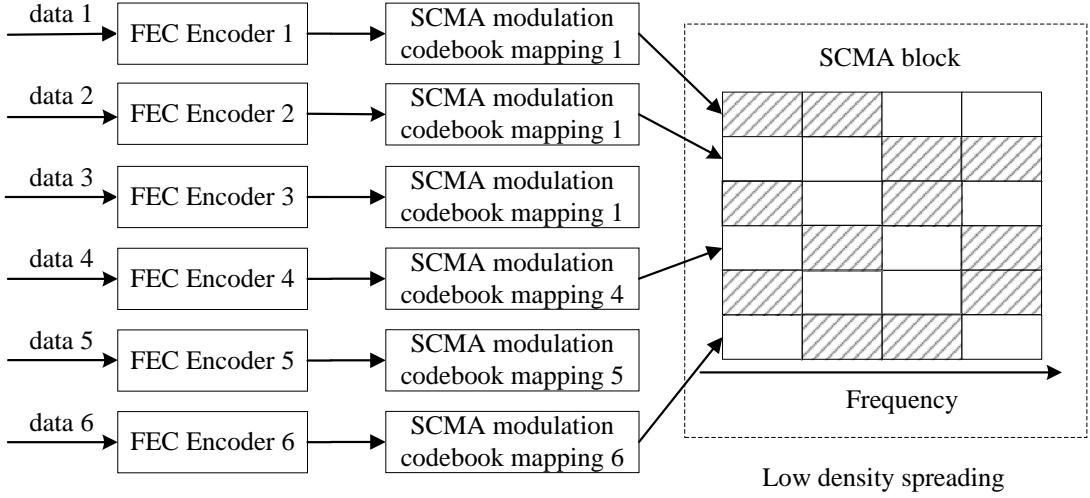}
	\caption{\small{Sparse code multiple access (SCMA).}}
	\label{fig:scma}
\end{figure}

In SCMA~\cite{SCMA}, by applying a sparse code book similar to the signature matrix in LDS, a certain number of resource blocks can support more users through spreading. Fig. \ref{fig:scma} demonstrates SCMA where six users share four resource blocks.
For the example in Fig. \ref{fig:scma}, the signature matrix can be expressed as
\begin{align}
\mathcal{\mathbf{S}} = \begin{bmatrix}
1 &0 &1 &0 &1 &0 \\
1 &0 &0 &1 &0 &1 \\
0 &1 &1 &0 &0 &1 \\
0 &1 &0 &1 &1 &0 \\
\end{bmatrix} \label{eq:scmasigmat}.
\end{align}
Although a part of the users share the same block,  another block would be adopted to distinguish different users when collisions occur.

Besides the sparse spreading,
SCMA utilizes multi-dimensional constellations to reduce the receiver complexity and further improve the SE. Attributed to the multi-dimension property, the constellation in one resource block can be projected into its subspace \cite{SCMAlp}. For example, a four-point QAM constellation can be projected to a three-point constellation. Even when two points  collide in one resource block or to say one dimension, they can be distinguished in the other used blocks. Due to fewer constellation points, the receiver complexity can be reduced. Moreover,  the constellation design  can focus on improving the detection performance. For example, a design based on constellation rotation and interleaving has been proposed in \cite{SCMArot}, which is able to achieve better BER performance compared to the simple LDS-OFDM.

Due to the sparse structure of the spreading matrix and the large minimum distance of the multi-dimensional constellation, the detection performance of SCMA becomes excellent even when the resource blocks are overloaded. At the receiver, MPA, which is usually adopted in the decoding of LDPC, is applied in the detection \cite{SCMArec,SCMAreclc}. Due to the sparsity, MPA could achieve near-optimal performance with a much lower complexity compared to the optimal maximum likelihood (ML) and the BCJR algorithms. However, the complexity is still relatively high for  user devices. Hence, SCMA also considers clustering the users based on the CSI and allocating different powers to different clusters. When the transmit powers among different clusters vary, the signals of different clusters can be detected by using SIC, which is similar to the power-domain NOMA. Within each cluster, different users can be distinguished by using MPA. As a result, the combination of SIC and MPA can reduce the complexity of the receiver significantly.




\subsection{NOMA Multiplexing in Multiple Domains}
Beyond multiplexing signals in the power or code domains, some of solutions for NOMA have been proposed to multiplex in multiple domains, such as the power domain, the code domain, and the spatial domain, in order to support massive connectivity for  5G networks. In Section III.A.2, we discussed multiple antenna based NOMA, where NOMA multiplexed in the power and spatial domains. We now  introduce another three types of typical NOMA schemes multiplexing in multiple domains: pattern division multiple access (PDMA)~\cite{PDMA}, building block sparse-constellation based orthogonal multiple access (BOMA), and lattice partition multiple access (LPMA)~\cite{LPMA}.


\subsubsection{PDMA}	\label{sec:PDMA}
In PDMA~\cite{PDMA},  non-orthogonal patterns are allocated to different users to perform  multiplexing. These patterns are carefully designed in the multiple domains of code, power, and space, to gain the SIC-amenable property. In the presence of this property, the low-complexity SIC based MPA multiuser detection method with reliable performance can be designed to run at the receiver side.


At the transmitter, similar to SCMA, the users in PDMA are also spread by a sparse signature matrix \cite{PDMA}. The main difference is that the number of resource blocks occupied by each user in PDMA can vary. For example, seven users can be multiplexed within three resource blocks through the following signature matrix
\begin{align}
\mathbf{S} = \begin{bmatrix}
1 &\sqrt{\frac{3}{2}} &\sqrt{\frac{3}{2}} &0 &\sqrt{3} &0 &0 \\
1 &\sqrt{\frac{3}{2}} &0 &\sqrt{\frac{3}{2}} &0 &\sqrt{3}  &0 \\
1 &0 &\sqrt{\frac{3}{2}} &\sqrt{\frac{3}{2}} &0 &0 &\sqrt{3} \\
\end{bmatrix}. \label{eq:pdmasigmat}
\end{align}
By utilizing the sparse signature matrix, PDMA can increase the system capacity through overloading. Moreover, users can also be multiplexed in other domains, such as power and space. In the same resource block, users can be distinguished by different powers as the power-domain NOMA or different precoders if MIMO is applied~\cite{PDMA2}.

At the receiver side, similar to SCMA, MPA can be adopted in detection due to the sparsity of the signature matrix. When  different clusters of users are multiplexed in power and space domains, MPA-SIC can be applied. The detection of users that are multiplexed in the same signature matrix is based on the MPA, which can provide excellent performance. Among different clusters in the power and space domains, SIC can be utilized to reduce the complexity. Besides, a turbo structure can be adopted to combine the detector with the decoder to further improve the performance~\cite{PDMAturbo}.

\subsubsection{BOMA}	\label{sec:BOMA}
\begin{figure}[!t]
	\centering
\includegraphics[width=6in]{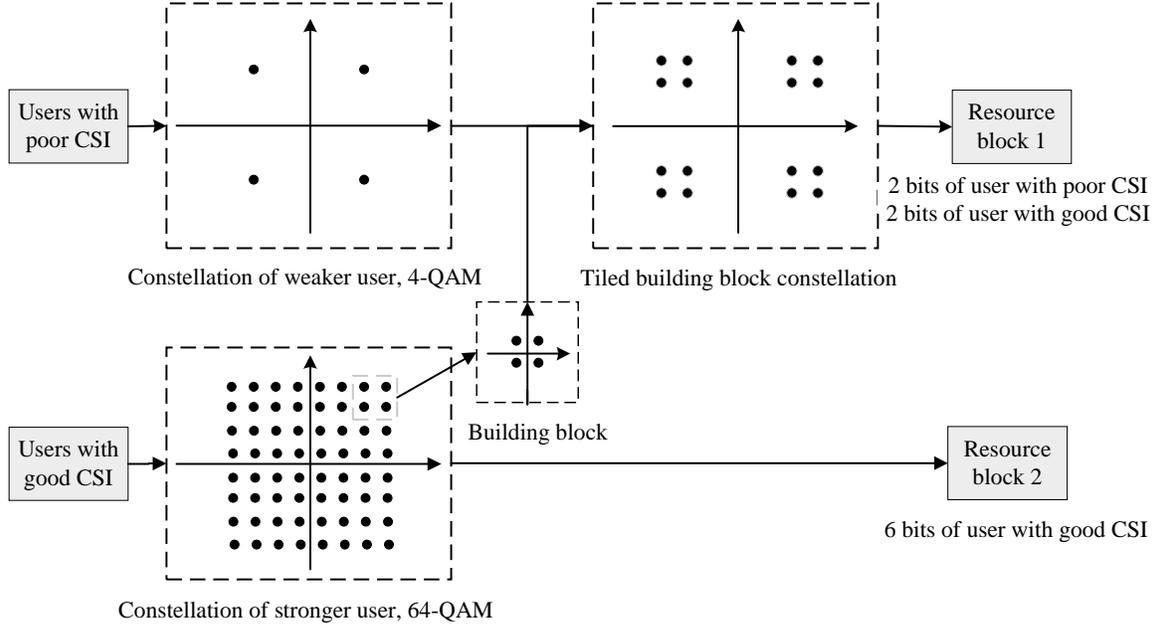}
	\caption{\small{Building block sparse-constellation based orthogonal multiple access (BOMA).}}
	\label{fig:boma}
\end{figure}

This technique  attaches the information from a user with good CSI to the symbols of a user with poor CSI. Thus the capacity of a multiuser system is increased significantly.

As shown in Fig.~\ref{fig:boma}, in order to achieve the same BER performance as a user with good CSI, the user with poor CSI should apply a coarse constellation  with a large minimum distance. Hence, the small building block that contains the data of the user with good CSI can be tiled in the constellation of the user with poor CSI~\cite{BOMA,BOMA_patent}. For the user with poor CSI, the center of the building block can be regarded as the constellation point and the tiled building block can be regarded as interference. When the size of the building block is much smaller than the minimum distance of the coarse constellation, the degradation of detection performance becomes minimal. Since the user with good CSI can detect the points in its own constellation, it can also detect all  points in the tiled building block constellation and decode the bits from itself.

The structure of BOMA is simple and similar to that adopted in current 4G systems. Only minor software changes are required so that BOMA can be easily implemented with the compatibility to massive MIMO, high frequency bands, and other requirements of 5G systems. Besides, BOMA  needs no complex power allocation and SIC receiver that are necessary for other NOMA schemes.

\subsubsection{LPMA}	\label{sec:LPMA}
In LPMA~\cite{LPMA}, the power domain and code domain are combined to multiplex users. Similar to power multiplexing in power-domain NOMA, the code in LPMA implements a multilevel lattice  that allocates different code levels to users with different CSI. Several types of codes can be adopted, such as Construction $\pi_A$ \cite{LPMAcodepiA} and Construction D \cite{LPMAcodeD}. For users with the poor CSI, the allocated codes have larger minimum distance  that can improve  detection performance. For users with better CSI, the allocated codes are with smaller minimum distance without degrading detection performance. At the receiver, a SIC decoder is adopted, which is similar to that found in power-domain NOMA.

Besides the code domain multiplexing, LPMA also adopts power multiplexing to enhance those users with poor CSI. With the aid of two degrees of freedom in the multiplexing, the design of LPMA becomes more flexible in comparison with power-domain NOMA. Even if a pair of users have similar CSI, they can still be multiplexed by adjusting the allocated code levels and power levels, therefore the complex user clustering mechanisms adopted in the power-domain NOMA schemes are not required in LPMA.

\subsection{Comparison}
In the above, we have introduced several typical NOMA schemes based on various multiplexing techniques. With different multiplexing techniques, the features of these NOMA schemes are different. The power-domain NOMA has a simple structure and can be easily combined with various technologies, such as MIMO and cooperative networks. However, user clustering and pairing  are required to get the user order in terms of CSI,  which increases the system complexity. Code-domain NOMA and the NOMA multiplexing in multiple domains, such as SCMA and PDMA, can obtain the spreading gain without requiring perfect CSI. Besides, the low-complexity near-optimal MPA detectors can achieve better performance than the SIC detectors. However, the coding also introduces redundancy, which degrades the system SE. Other NOMA schemes also have their specific features. For example, BOMA has a simple structure and can be easily compatible to the current LTE systems while user pairing  is also required, which reduces the flexibility. LPMA can achieve multilevel in both the power domain and the code domain, which saves the cost of user clustering. However, the multilevel code requires corresponding non-binary or nested binary channel coding. The key techniques and features of the aforementioned NOMA schemes are summarized in Table.~\ref{table:comparison}. It is noted that each NOMA scheme has its advantages and limitations, which fits different application situations. Actually, it is an adaptive configuration to realize the trade-off between performance and implementation complexity. For example, if there exists a large difference among users' channel conditions due to the near-far effect or in moving networks, power-domain NOMA with a SIC receiver can be used with relatively low complexity. On the other hand, if the application scenarios require high reliability, especially when channel condition is bad or the location distribution of users is concentrated, SCMA is a feasible solution due to its shaping gain and near-optimal MPA detection.
\newcommand{\tabincell}[2]{\begin{tabular}{@{}#1@{}}#2\end{tabular}}
\begin{table}[!t]
	\caption{\small{Comparisons of NOMA schemes}}
	\begin{center}
		\centering
		\begin{tabular}{|l|l|l|l|}
			\hline
			\centering
			\textbf{Schemes}& \textbf{Characteristics} & \textbf{Advantages} &\textbf{Disadvantages} \\
			\hline
			\centering
			\tabincell{l}{Power-domain\\ NOMA} & Power domain multiplexing &\tabincell{l}{High SE\\ Compatible to other techniques} & \tabincell{l}{Need user pairing\\ Error propagation in SIC}   \\
			\hline
			\centering
			LDS-CDMA & \tabincell{l}{Sparse spreading\\ CDMA}& \tabincell{l}{No need of CSI\\ Near-optimal MPA detector} &\tabincell{l}{Redundancy from coding} \\
			\hline
			\centering			
			LDS-OFDM & \tabincell{l}{Sparse spreading\\ OFDM} & \tabincell{l}{No need of CSI\\ Near-optimal MPA detector\\ More fit for wideband than LDS-CDMA} &\tabincell{l}{Redundancy from coding} \\
			\hline
			\centering	
			SCMA & \tabincell{l}{Sparse spreading\\ Multi-dimensional constellation} & \tabincell{l}{No need of CSI\\ Near-optimal MPA detector\\ More diversity than simple LDS } &\tabincell{l}{Redundancy from coding\\ Difficult to design optimal codebook} \\
			\hline
			\centering
			PDMA & \tabincell{l}{Sparse spreading\\ Multiplexing in power, code, \\ and spatial domains} &\tabincell{l}{ More diversity\\ Near-optimal MPA detector\\ Low-complexity receiver} &\tabincell{l}{Redundancy from coding\\ Difficult to design optimal patterns}  \\
			\hline
			\centering
			BOMA & Tiled building block &\tabincell{l}{Simple structure\\ Compatible to current system \\ Low-complexity receiver} &\tabincell{l}{Need user pairing\\ Not very flexible} \\
			\hline
			\centering
			LPMA & \tabincell{l}{Multilevel lattice code \\ Multiplexing in power\\ and code domains } & No need for user clustering &Specific channel coding \\
			\hline
		\end{tabular}
	\end{center}
	\label{table:comparison}
\end{table}


\subsection{Future Work}	\label{sec:futureworkacess}
Several NOMA  schemes have been discussed in this section. Even if using different  techniques, these schemes share the same spirit to utilize non-orthogonality to increase the system capacity and support more users by the limited resource blocks. Beyond the existing work, more research is necessary to improve the performance of these NOMA schemes from the following aspects.

The MPA-SIC detection method is usually applied in SCMA and PDMA, in which the user clustering mechanism affects the performance of the method significantly. When users are asynchronous, those with similar time delays should be divided into the same cluster for better performance. If the delays vary a lot among the users within the same cluster, interference among different users becomes large and may break the sparse structure. Multi-branch technique~\cite{multibranch} can be applied to improve the performance by  regarding each cluster as a branch. By calculating each branch in parallel and selecting the best result as the final one, the performance could be improved compared to  the single clustering approach.

The joint design of new modulation and NOMA schemes is an important direction to be explored in  5G networks. Some of the NOMA schemes, especially the LDS based code-domain NOMA, are  based on OFDM, where the output of the sparse spreading matrix is mapped into orthogonal subcarriers. In general, how to properly combine the modulation and NOMA scheme is under research. For example, for the combination of SCMA and f-OFDM, the short CP of f-OFDM could introduce ISI and ICI when the subband is narrow and degrade the detection performance of SCMA. If the RISIC algorithm is adopted to cancel the interference introduced by f-OFDM, the multiuser detection of SCMA should be included in the iteration of CP reconstruction, which poses a requirement of joint design approaches for the receivers. 

The design of modulation and MA schemes for high frequency bands (above 40 GHz) is beginning to receive increased iterest. The millimeter-wave (mmWave) and
Terahertz (THz) bands appear to be good candidates to decrease spectrum sacristy due to the availabilities in current circuit design~\cite{Li:mmwave:2016,mmwavesurvey}. However, the propagation properties of mmWave and THz bands have shown to be quite poor, which brings new challenges on system designs. For example, noise is the major limitation of mmWave and THz bands, which makes the transmit power levels extremely important and ultimately impacts the classes of applications that can use them (e.g. IoT). Moreover, high level impairments including carrier frequency offset (CFO) and phase noise also need to be considered in mmWave  and THz bands as they are  noise-limited. Nevertheless, there is already a study on  NOMA based mmWave communications~\cite{Ding:mmwave:NOMA}, we may further see analyses of such systems based on practical scenarios in the future.

\section{Conclusions} \label{sec:conclusion}
In this article, we provide a comprehensive survey covering the major promising candidates for modulation and multiple access (MA) in fifth generation (5G) networks. From our discussion, we can see that new modulations for orthogonal MA can be adopted to reduce out-of-band leakage while meeting the diverse demands of 5G networks. Non-orthogonal MA is another promising approach that marks a deviation from the previous generations of wireless networks. By utilizing non-orthogonality, we have convincingly shown that 5G networks will be able to provide enhanced throughput and massive connectivity with improved spectral efficiency.

%








\end{document}